%% file: main.tex
\documentclass[%
reprint,
amsmath,amssymb,
aps,]{revtex4-2}

\usepackage{graphicx}
\usepackage{dcolumn}
\usepackage{bm}
\usepackage{revquantum}
\usepackage{placeins}
\usepackage[caption=false]{subfig}
\usepackage{makecell}

\begin{document}


\title{Quantum error correction in the NISQ regime for sequential quantum computing}


\author{Arvid Rolander}

\author{Adam Kinos}



\author{Andreas Walther}
\email[]{andreas.walther@fysik.lth.se}
\affiliation{Department of Physics, Lund University, SE-22100 Lund, Sweden}


\date{\today}

\begin{abstract}
\input{Sections/Abstract}
\end{abstract}


\maketitle

\section{\label{sec:Introduction}Introduction}
\input{Sections/Introduction}

\section{\label{Sec:SimulatedCodes}Simulated Codes}

\input{Sections/IntroOnCodesAndSims}

\section{\label{sec:PseudoThresh}Pseudothresholds}
\input{Sections/Pseudothresholds}

\section{\label{sec:pseudopara}Pseudo-Parallelism}
\input{Sections/Pseudoparallelism}
\section{\label{sec:gain}Comparison with Physical Circuit}
\input{Sections/Gain}

\section{\label{sec:conclusions}Conclusions}
\input{Sections/Conclusions}

\begin{acknowledgments}
We acknowledge helpful discussions and comments from Stefan Kröll, Lars Rippe, and Klaus Mølmer. The research leading to these results has received funding from the European Union's Horizon 2020 research and innovation programme under grant agreement No 820391 (SQUARE), as well as from the Swedish Research Council (grant No 2015-03989).
\end{acknowledgments}

\clearpage
\appendix
\section{\label{sec:bckgrnd}Background \& Theory}
\input{Sections/Background_Theory}

\section{\label{sec:DMSim}Density Matrix Simulation}
\input{Sections/DensityMatrixSim}


\bibliography{bibliography}

\end{document}

%% file: Sections/Abstract.tex
We use density matrix simulations to study the performance of three distance three quantum error correcting codes in the context of the rare-earth-ion-doped crystal (RE) platform for quantum computing. We analyze pseudothresholds for these codes when parallel operations are not available, and examine the behavior both with and without resting errors. In RE systems, resting errors can be mitigated by extending the system's ground state coherence time. For the codes we study, we find that if the ground state coherence time is roughly 100 times larger than the excited state coherence time, resting errors become small enough to be negligible compared to other error sources. This leads us to the conclusion that beneficial QEC could be achieved in the RE system with the expected gate fidelities available in the NISQ regime. However, for codes using more qubits and operations, a factor of more than 100 would be required. Furthermore, we investigate how often QEC should be performed in a circuit. We find that for early experiments in RE systems, the minimal $\llbracket5,1,3\rrbracket$ would be most suitable as it has a high threshold error and uses few qubits. However, when more qubits are available the $\llbracket9,1,3\rrbracket$ surface code might be a better option due to its higher circuit performance. Our findings are important for steering experiments to an efficient path for realizing beneficial quantum error correcting codes in early RE systems where resources are limited.

%% file: Sections/Introduction.tex
Several platforms have been proposed for building scalable quantum computers, such as superconducting qubits and trapped ions. A promising platform, and the focus of this work, are rare-earth ion doped crystals \cite{WesenbergMolmer2003,REQC_Roadmap}. These have some particularly attractive features, such as ground state life times of days \cite{Konz2003} and coherence times of hours \cite{Zhong2015}, high qubit connectivity \cite{AKinos2021QP} which enables entangling operations on non-nearest neighbor qubits, and high spatial density, enabling efficient integration with optics. Rare earth (RE) systems also have some unique challenges, the main one relevant to this work being a need to perform operations sequentially to minimize crosstalk between qubit ions. The RE system uses two hyperfine groundstates as the $\ket{0}$ and $\ket{1}$ states, and a mediant excited state $\ket{e}$ used during gate operations. The lifetime $T_1$ and coherence time $T_2$ are longer for the ground states than the excited states \cite{Equall1994,Arcangeli2014,Konz2003,Zhong2015,Alexander2007}. 

An inherent weakness of quantum computing is its sensitivity to noise, and Quantum Error Correction (QEC) is one of the proposed solutions. Efforts have been made to study QEC in the specific context of ion traps \cite{Bermudez2017, Debroy2020, Trout2018}, superconducting qubits \cite{OBrien2017,Ghosh2012} and systems based on nitrogen-vacancy defects in diamond \cite{Waldherr2014}, but is also being discussed for ensemble qubits in stoichiometric RE-doped crystals \cite{Ahlefeldt2020}. Because even resting qubits can accumulate errors, e.g. through $T_2$ decoherence, some degree of parallelism of operations is seen as a requirement for beneficial QEC \cite{Steane1998, gottesman2009introduction} \cite[pp. 482]{nielsen2010quantum}. However, because the ground state lifetime of the rare-earths can be extended to several orders of magnitude longer than the time for a typical gate operation, this could potentially be relaxed in the RE case. A key point of investigation in this work is therefore whether beneficial quantum error correction can be achieved using sequential operations, provided that the ground state coherence time is long enough compared to the excited state coherence time. By extending the ground state coherence time enough, a sort of pseudo-parallelism could possibly be achieved, in the sense that the resting errors are small enough compared to other error sources to be neglected. We also investigate how well the different quantum error correction protocols could be expected to perform in a rare earth system when performing an algorithm.  

We have chosen to focus on QEC in the Noisy Intermediate-Scale Quantum (NISQ) regime \cite{Preskill2018}. This includes devices using 50-100 noisy qubits, and could be available in the near future \cite{REQC_Roadmap}. In this regime, arbitrarily high fidelity cannot be achieved through concatenated codes or large surface code lattices due to lack of resources in terms of available qubits and lower gate fidelities. Instead, it becomes important to use QEC codes with high threshold errors and using few qubits. It is still a challenge for many platforms to demonstrate beneficial QEC experimentally. An important goal is therefore to investigate what would be required of a RE system in order to experimentally demonstrate a gain from using QEC. 

To summarize, the questions we want to answer are:\begin{itemize}
    \itemsep-1em
    \item Could beneficial QEC be demonstrated in the RE system  using current protocols with projected fidelities?
    \item Can an extended resting $T_2$ be sufficient to overcome the demands for parallelism?
    \item How much can be gained from QEC, given reasonable parameter values?
\end{itemize}

%% file: Sections/IntroOnCodesAndSims.tex
To answer the questions posed in Sec. \ref{sec:Introduction}, we have used density matrix simulations (see Appendix \ref{sec:DMSim}) to investigate three distance three (\cite{nielsen2010quantum}, see appendix. \ref{ssec:QEC}) QEC codes in the context of rare earth quantum computing. Distance three codes have logical codewords that are separate from each other in three places, and is the minimum distance needed to correct for any one arbitrary error. The density matrix approach was chosen for its ability to model arbitrary quantum processes. The QEC codes we investigate are the $\llbracket5,1,3\rrbracket$ code \cite{Laflamme1996PerfectCode} using the flag syndrome extraction scheme of \cite{Chao2018}, the $\llbracket7,1,3\rrbracket$ Steane code \cite{Steane1996,nielsen2010quantum} using the flag syndrome extraction scheme of \cite{Chamberland2018flagfaulttolerant} and the $\llbracket9,1,3\rrbracket$ Surface-17 as described in \cite{Tomita2014Surface17}, using the lookup table decoder of \cite{Tomita2014Surface17} with 3 rounds of stabilizer measurements. Because projective measurements directly on the data qubits destroys the stored information, ancillary qubits have to be used for readout. An important aspect when considering QEC in the NISQ-regime is the efficient use of qubits as a resource, due to the limited number of qubits available. For this reason, the flag protocols of \cite{Chao2018, Chamberland2018flagfaulttolerant} are very appealing, as only two ancilla qubits are required. This can be compared to the Shor-syndrome extraction method, which would need four ancillas \cite{shor1996faulttolerant} or the Steane protocol which requires at least seven ancillas for the Steane code \cite{Steane97, Steane2004}. The main feature which makes the surface code attractive is its high threshold error, which could be as high as 1\% \cite{Fowler2012PRL}. Surface-17 uses 17 qubits in total, of which nine are data qubits and eight ancillas used for measurements. However, the high qubit connectivity of the RE system makes it so that we are not limited to nearest neighbor interactions for entanglement and parity measurements \cite{AKinos2021QP}. In addition to this, since we propose using fully sequential operations, we could reduce the number of ancillas required to just one, using only 9+1 qubits in total for code we from here onward call Surface-9(1). This comes at the cost of increased run time for the error correction scheme. Table \ref{tab:qubitReqs} shows the qubit requirements for each of the simulated codes with different readout schemes.

\begin{table}[h]
    \centering
    \caption{\label{tab:qubitReqs}Overview of the QEC codes that were simulated including the number of qubits used to represent the logical state plus the number of ancilla qubits}
    \begin{tabular}{|r|c|c|} \hline
      \thead{Code:}   & \thead{Readout Scheme/\\Decoder:} & \thead{No. Qubits \\+ Ancillas:}\\ \hline
       $\llbracket5,1,3\rrbracket$  & Flag &    5+2 \\ \hline
       Steane $\llbracket7,1,3\rrbracket$  & Flag & 7+2 \\ \hline
       Surface-9(1) $\llbracket9,1,3\rrbracket$  & \makecell{Lookup table, \\ 3 measurement \\rounds} & \makecell{9+1 \\ (9+8)} \\ \hline
    \end{tabular}

\end{table}


%% file: Sections/Pseudothresholds.tex
In this section we present simulation results intended to investigate the pseudothresholds \cite{DBLP:journals/qic/SvoreCCA06} of our chosen QEC codes and extraction schemes. The pseudothreshold is the break-even point for a given QEC protocol, and can be seen as the point where the logical error rate is equal to the physical error rate. The simulations were performed by applying one round of imperfect error correction, and the logical error rate was calculated as the average over all starting states corresponding to the six axes of the Bloch sphere i.e. $\ket{0_L}, \ \ket{1_L}, \ \ket{+_L}, \ \ket{-_L}$ etc. For more information see appendix \ref{sec:PerformanceMeasure}. 

We define the two qubit gate (TQG) error rate $p_\mathrm{TQG}$ as the main error parameter, and error rates for other operations are defined in terms of this. The single qubit gate (SQG) error rate $p_\mathrm{SQG}$ was set to $p_\mathrm{TQG}/10$, which is reasonable for RE systems \cite{kinos2021designing}. We choose to treat qubit initialization and readout as single qubit operations, and therefore the  initialization and readout error rates, $p_\mathrm{init}$ and $p_\mathrm{RO}$, were set to two thirds of the SQG rate. Two thirds is used because we choose to always perform initialization and readout in the $\left\{\ket{0},\ket{1}\right\}$ basis, and they are therefore unaffected by phase errors. A summary of the values used is given in table \ref{tab:SatT2_params}, together with the optical coherence time $T_\mathrm{2, optical}$ and typical duration times for the different operations. The values of the operations times were chosen to resemble the ones found in \cite{kinos2021designing} and \cite{Walther2015}. In order to reach an initialization duration shorter than the excited state lifetime a technique based on stimulated emission to another short lived crystal field level can be used \cite{Lauritzen2008}. For a thorough description of the error models used, see appendix \ref{sec:errormodel}. 
\begin{table*}
    \caption{Parameter values used in simulations.}
    \label{tab:SatT2_params}
    \begin{tabular}{|c|c|c|c|c|c|}\hline 
      \thead{Parameter:} & \thead{Duration ($t_{dur}$):} & \thead{Error Symbol:}  & \thead{Error Value:} & \thead{Resting Error \\(figure \ref{fig:pseudothreshold resterr}):}& \thead{Resting Error \\(figure \ref{fig:SaturationT20}):}  \\\hline
      
      Single qubit gate & 5 $\mu$s & $p_\mathrm{SQG}$ & $p_\mathrm{TQG}$/10 & $p_\mathrm{TQG}$ & $\gamma = e^{\frac{-t_{dur}}{2T_\mathrm{2,spin}}}$\\ \hline
      
      Two qubit gate & 10 $\mu$s & $p_\mathrm{TQG}$ & $p_\mathrm{TQG}$ & 2$p_\mathrm{TQG}$ & $\gamma = e^{\frac{-t_{dur}}{2T_\mathrm{2,spin}}}$ \\ \hline
      
      Readout  & 100$\mu$s & $p_{\mathrm{RO}}$ & $ p_\mathrm{TQG} \cdot 2/30$ & $20p_\mathrm{TQG}$ & $\gamma = e^{\frac{-t_{dur}}{2T_\mathrm{2,spin}}}$\\\hline
      
      Initialisation & 10 $\mu$s & $p_{\mathrm{init}}$ & $ p_\mathrm{TQG} \cdot 2/30$ & 2$p_\mathrm{TQG}$ & $\gamma = e^{\frac{-t_{dur}}{2T_\mathrm{2,spin}}}$ \\\hline
        
    \end{tabular}
\end{table*}
In the case were a resting error is included, we assume that the damping coefficient $\gamma$ is $2p_\mathrm{TQG}$ for TQG and initialization, $p_\mathrm{TQG}$ for SQG and $20p_\mathrm{TQG}$ for readout. In other words, we assume that the resting errors reduce as the two qubit error reduces. The relative ratios between the different resting errors come directly from the ratios of their durations.   

\begin{figure*}
        \subfloat[\label{fig:pseudothreshold resterr}]{
        
        \includegraphics[width=0.45\textwidth]{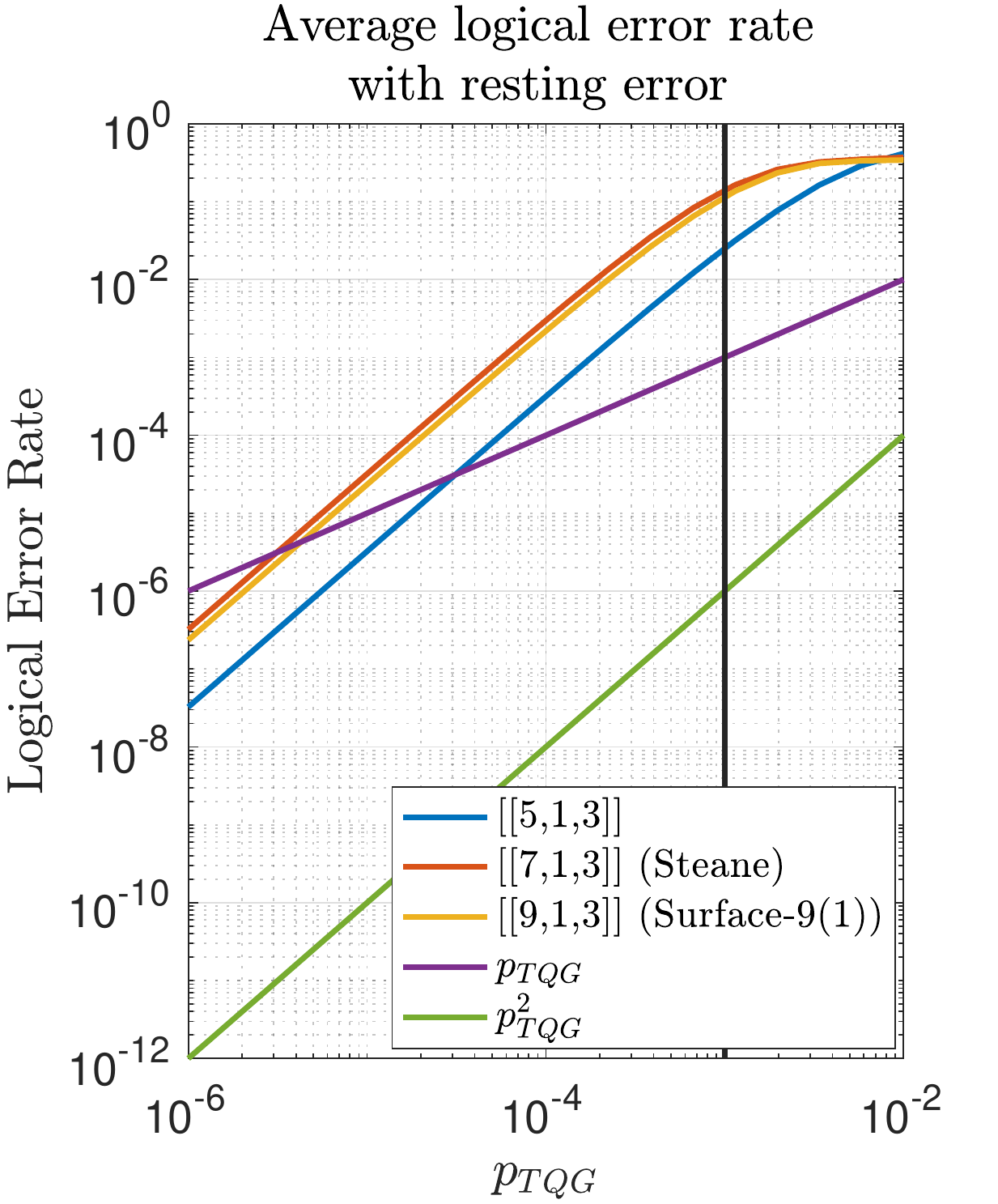}

    }
    \hfill
    \subfloat[\label{fig:pseudothresholdnoresterr}]{
        
        \includegraphics[width=0.45\textwidth]{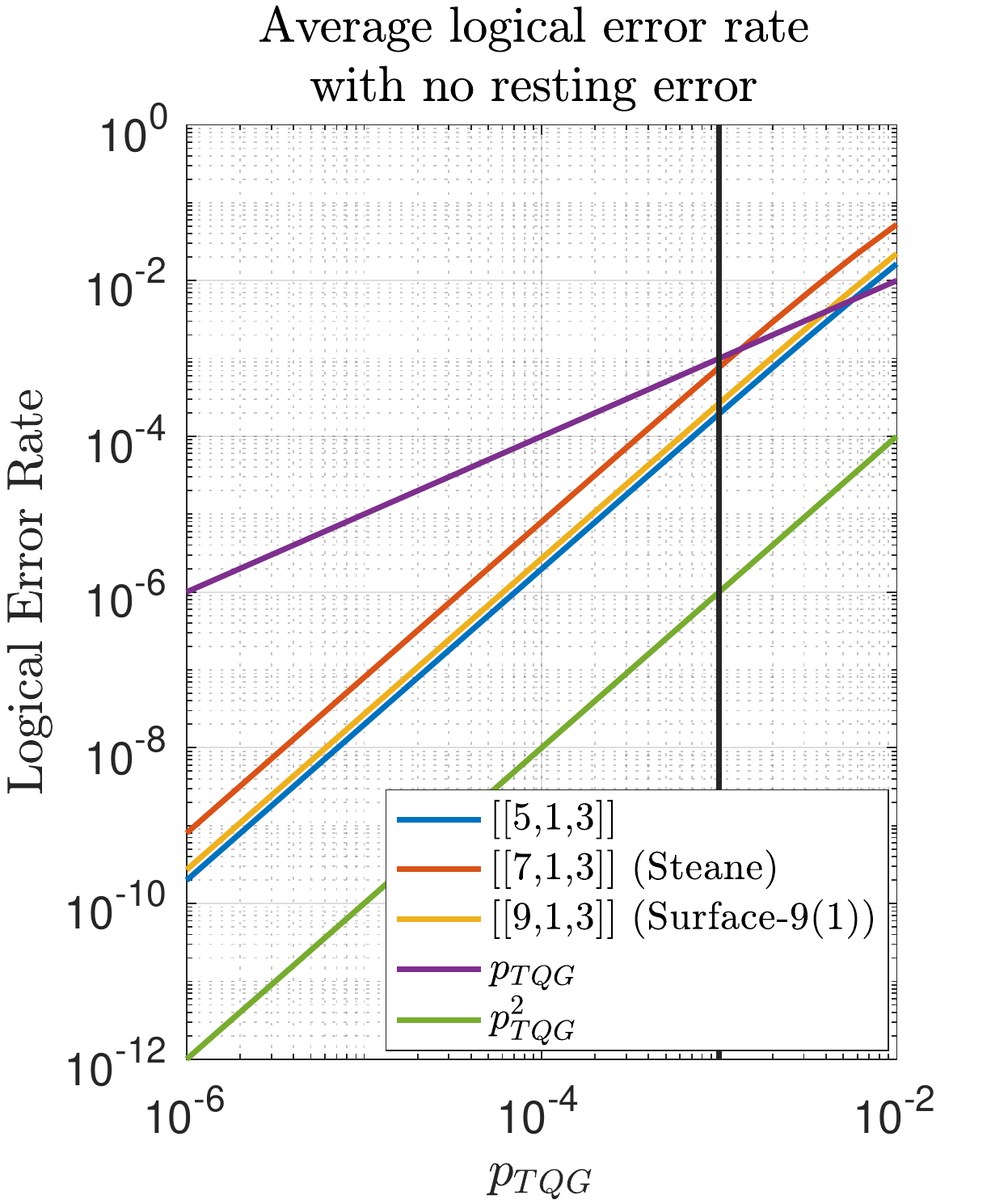}

     }

   \caption{\label{fig:pseudothreshold} Pseudothresholds for our chosen codes with resting errors either included (\ref{fig:pseudothreshold resterr}) or excluded (\ref{fig:pseudothresholdnoresterr}). The logical error rates were calculated as the average for starting states corresponding to the six axes of the Bloch sphere. In figure \ref{fig:pseudothreshold resterr}, we have assumed that the resting errors for the different operations scale with $p_\mathrm{TQG}$. This assumes that if we are able to improve $p_\mathrm{TQG}$, we will be able to improve the resting errors by the same factor. Black, vertical lines mark the points where the physical two-qubit error rate is $p_\mathrm{TQG}=10^{-3}$. This is approximately the error rate where a physical system, with today's protocols \cite{REQC_Roadmap}, could be expected to lie \cite{kinos2021designing}.}
    
\end{figure*}

Figure \ref{fig:pseudothreshold resterr} shows how the logical error rate varies as a function of $p_\mathrm{TQG}$, i.e. the two qubit error rate, when all error sources are included, and Figure \ref{fig:pseudothresholdnoresterr} shows the case when resting errors are excluded. The case with no resting errors is relevant since the ground state coherence time of RE systems is much longer then the gate coherence time. Black, vertical lines mark the points where the physical TQG error rate is $p_\mathrm{TQG} = 10^{-3}$. 
Interestingly, the $\llbracket5,1,3\rrbracket$, not Surface-9(1), performs best both when resting errors are included and excluded. There are several reasons for this, but the main one seems to be the comparatively low SQG error that we have used. Lowering the SQG error compared to the TQG error improves the performance of all codes, but the effect is larger for the $\llbracket5,1,3\rrbracket$ code than the other two. This discrepancy can be explained by looking at what proportion of the operations used in each protocol are single qubit operations: the $\llbracket5,1,3\rrbracket$ uses 60\% single qubit operations, Surface-9(1) uses 50\% and the Steane code uses between 50\% and 62.5\% depending on the syndrome measurement results. Another reason for the lower surface code performance could be the choice of  the lookup table decoder, as other decoders could yield better results \cite{OBrien2017}. Figure \ref{fig:pseudothreshold} also shows that the surface code and Steane code are more affected by resting errors than the $\llbracket5,1,3\rrbracket$ code. This can be explained by looking at the number of operations and qubits used in each of the protocols; The $\llbracket5,1,3\rrbracket$ uses 55 operations in the best case and 100 operations in the worst case scenario, and the Steane code uses 96 operations in the best case and 276 operations in the worst case. Surface-9(1) always uses 144 operations. The different cases arise as the flag protocols follow different paths depending on if the circuits flag or not. Since all operations are performed sequentially, this means that Surface-9(1) and the Steane code have more idle time where resting errors are accumulated on more qubits compared to the $\llbracket5,1,3\rrbracket$ code. The higher number of qubits used by the Steane code and Surface-9(1) compared to the $\llbracket5,1,3\rrbracket$ code also contributes to the higher error, as they have more qubits in waiting for every operation.

%% file: Sections/Pseudoparallelism.tex
In this section we present results aiming to investigate pseudo-parallelism by varying the magnitude of resting errors. The simulations were again performed by applying one round of imperfect error correction, followed by calculating the error rate as the average over the six axes of the Bloch sphere. For these simulations, we used a different model for resting errors to that of Sec. \ref{sec:PseudoThresh}. Instead of assuming the resting errors follow $p_\mathrm{TQG}$, we kept the TQG error rate fixed at $p_\mathrm{TQG} = 10^{-3}$, but assumed that we can improve the resting errors by increasing the ground state coherence time $T_\mathrm{2, spin}$. Therefore, the magnitudes of the resting errors were calculated as  $\gamma = 1-e^{\frac{-t_\mathrm{dur}}{2T_\mathrm{2, spin}}}$ (see appendix \ref{ssec:Idleerrors}) with the duration times found in Table \ref{tab:SatT2_params}.

\begin{figure}[ht]
    \includegraphics[width=0.5\textwidth]{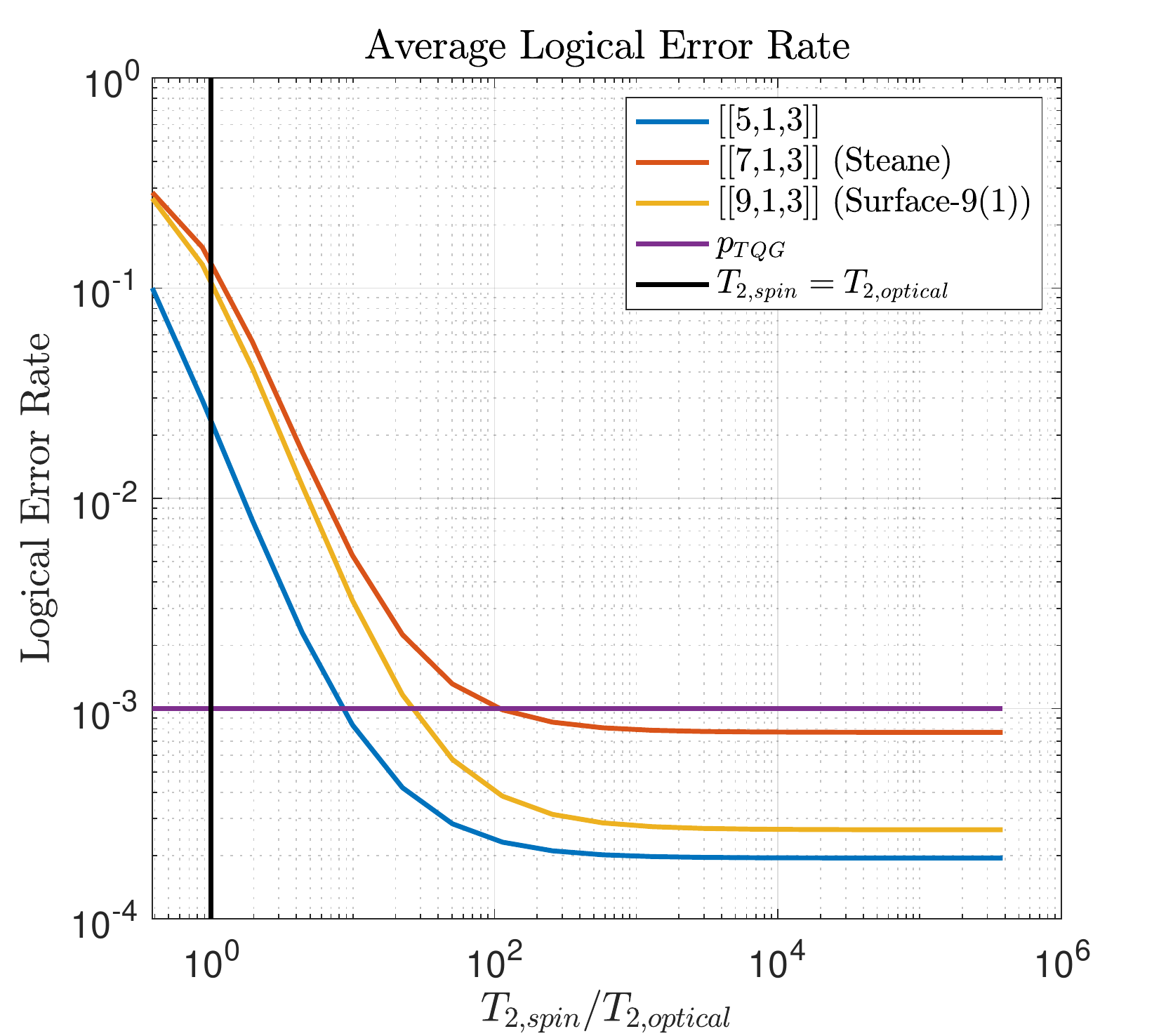}
    
    \caption{\label{fig:SaturationT20}Logical error rate as a function of the ground state coherence time $T_{2, \mathrm{spin}}$ which varies between 1 ms and 1 hour. The x-axis is scaled by the excited state coherence time $T_{2, \mathrm{optical}} = 2.5$ ms. In this simulation we have used a different model of the resting error compared to the one used in figure \ref{fig:pseudothreshold resterr}. Here, we assume a constant value of the two qubit error rate, $p_\mathrm{TQG} = 10^{-3}$, but the resting errors still change due to changes in $T_\mathrm{2,spin}$, see Table \ref{tab:SatT2_params}. This model therefore assumes that it is possible to reduce resting errors even though the two qubit error rate is kept constant. The models give the same values for all errors when $p_\mathrm{TQG}=10^{-3}$ and $T_\mathrm{2,spin} = 2.5 $ ms, which are reasonable values \cite{kinos2021designing} for a RE system using current protocols \cite{REQC_Roadmap}. These points are marked with black lines in both figure \ref{fig:pseudothreshold resterr} and figure \ref{fig:SaturationT20}. Furthermore, in a RE system it is possible to reach values for $T_\mathrm{2,spin}$ of hours \cite{Zhong2015}. This corresponds to points to the far right in figure \ref{fig:SaturationT20}, where the resting errors are small enough compared to other error sources to be neglected. The errors here have approximately the same values as in figure \ref{fig:pseudothresholdnoresterr} when $p_\mathrm{TQG}=10^{-3}$, which is marked with a black line in figure \ref{fig:pseudothresholdnoresterr}. This shows that the resting errors can be reduced enough to become negligible.}


\end{figure}

The simulation results are shown in figure \ref{fig:SaturationT20}, where we have chosen to scale the ground state coherence time $T_\mathrm{2, spin}$ by the excited state coherence time $T_\mathrm{2, opt}=2.5$ ms. 
The point where $T_\mathrm{2, spin}/T_\mathrm{2, optical} = 1$ is marked with a black, vertical line. At this point, the error rate and magnitude of the resting errors are approximately the same as in figure \ref{fig:pseudothreshold resterr} at the vertical, black line where $p_\mathrm{TQG}=10^{-3}$. As the value of $T_\mathrm{2, Spin}/T_\mathrm{2, Optical}$ increases, a plateau is reached where the resting errors are insignificant, and the logical error rate is the same as at the vertical black line in figure \ref{fig:pseudothresholdnoresterr}, where resting errors are not included. This shows that even under the constraint that all operations must be performed sequentially, resting errors can be reduced to the point of being negligible in comparison with other error sources.

In a RE system the $T_2$ can be improved by various actions, such as applying a magnetic field in the right direction \cite{Zhong2015} or applying dynamic decoupling sequences \cite{Souza2011}. However, these actions have a cost; for instance, applying a magnetic field would also split the atomic levels making the qubit space more complicated, and running dynamic decoupling sequences would make the gate sequences more involved. For these reasons, it is interesting to see that there is is a level beyond which no further gain to the resting $T_2$ can be obtained. This allows experimentalists to hit a specific target rather than arbitrarily increasing the spin $T_2$, which serves to minimize the experimental efforts required to realize a quantum computer. 

It is important to note that such a target spin $T_2$ will change for higher distance codes, but we can estimate by how much: Each time we perform an operation on one of the data qubits, we assume that there is a probability in the order of $p_\mathrm{TQG}$ for a gate error to occur. If we view resting errors as $Z$ errors (see App. \ref{ssec:Idleerrors}) we can say that the probability for a single resting error to occur on just one of the qubits is roughly $(N-1)\cdot p_\mathrm{rest}$, where $N$ is the number of qubits, $p_\mathrm{rest}$ the probability for a resting error on a single qubit, and we assume that higher orders of $p_\mathrm{rest}$ are negligible. To be able to neglect resting errors, the probability for them occurring must be significantly smaller than the probability of a gate error. Studying figure \ref{fig:SaturationT20}, we see that the error rates flatten at roughly $T_\mathrm{2,spin} = 100 T_\mathrm{2,opt}$. Since the codes we use here use approximately 10 qubits, this means that resting errors become negligible approximately when the probability of only one resting error occurring on any of the qubits is an order of magnitude lower than the probability of a gate error, which seems reasonable. 

In the NISQ regime, we aim for having roughly 100 total qubits. Thus, implementing a distance 11 Surface code might be a reasonable future goal. The distance 11 Surface code uses $11^2=121$ data qubits, which, by our argument, would require a $T_\mathrm{2,spin}$ that is roughly 10 times longer than that required for Surface-9(1), i.e. $T_\mathrm{2,spin} = 1000 T_\mathrm{2,opt}$. With $T_\mathrm{2,opt}=2.5$ ms this means the distance 11 code would require $T_\mathrm{2,spin} = 2.5$ s. Achieving a spin $T_2$ of this magnitude is still very realistic in RE systems, and we therefore see reason to be optimistic. Although it can be noted that the method used to increase the T2 might have consequences, such as splitting levels or complicating sequences, and it remains for a future studies to determine the most optimal way to accomplish this. It is also important to note that a typical RE quantum processing node will likely have around 100 qubits, and operations in different nodes will be able to be performed in parallel \cite{REQC_Roadmap}. This means that as long as we can achieve a spin $T_2$ that is long enough for 100 qubits, scaling up further will be possible. 

%% file: Sections/Gain.tex
In this section we present results meant to study the performance of the error correcting codes when used in a circuit. From the results of section \ref{sec:pseudopara} we can expect that it is experimentally feasible to be able to reach a point where the ground state coherence time is not the main limitation when performing error correction. For this reason, we will omit ground state dephasing in the following. The simulated experiment consists of applying a logical NOT gate ($X_L$ if the state is $\ket{0_L}$, $Z_L$ for $\ket{+_L}$ etc.) $n_g$ times, which for distance three codes corresponds to 3 physical Pauli gates, followed by a round of error correction. The logical error rate is then calculated as the average for starting states corresponding to the six axes of the Bloch sphere as before. We again used the ratios for the physical error rates given in table \ref{tab:SatT2_params}.

To be able to compare the performance of the logical circuit to the equivalent physical circuit without error correction, we define a gain parameter as the ratio of the total error of the physical circuit and the total error of the logical circuit. This parameter tells us by which factor the error per gate changes when error correction is used compared to the physical circuit without error correction. At this point, it seems reasonable to ask how the gain should behave as a function of $n_g$. From a more holistic perspective, it could be argued that higher values of $n_g$ should give higher gain since error correction has considerable overhead in number of operations. For small values of $n_g$ the contribution of this overhead to the total error could be large in comparison to the contribution from increasing $n_g$. However, for large values of $n_g$ both the logical and physical qubit will approach a completely mixed state, and the gain should equal one as there will be no way of guessing the correct state. Therefore, we would expect to find an optimal value. 
\begin{figure}[ht]
    \includegraphics[width = 0.5\textwidth]{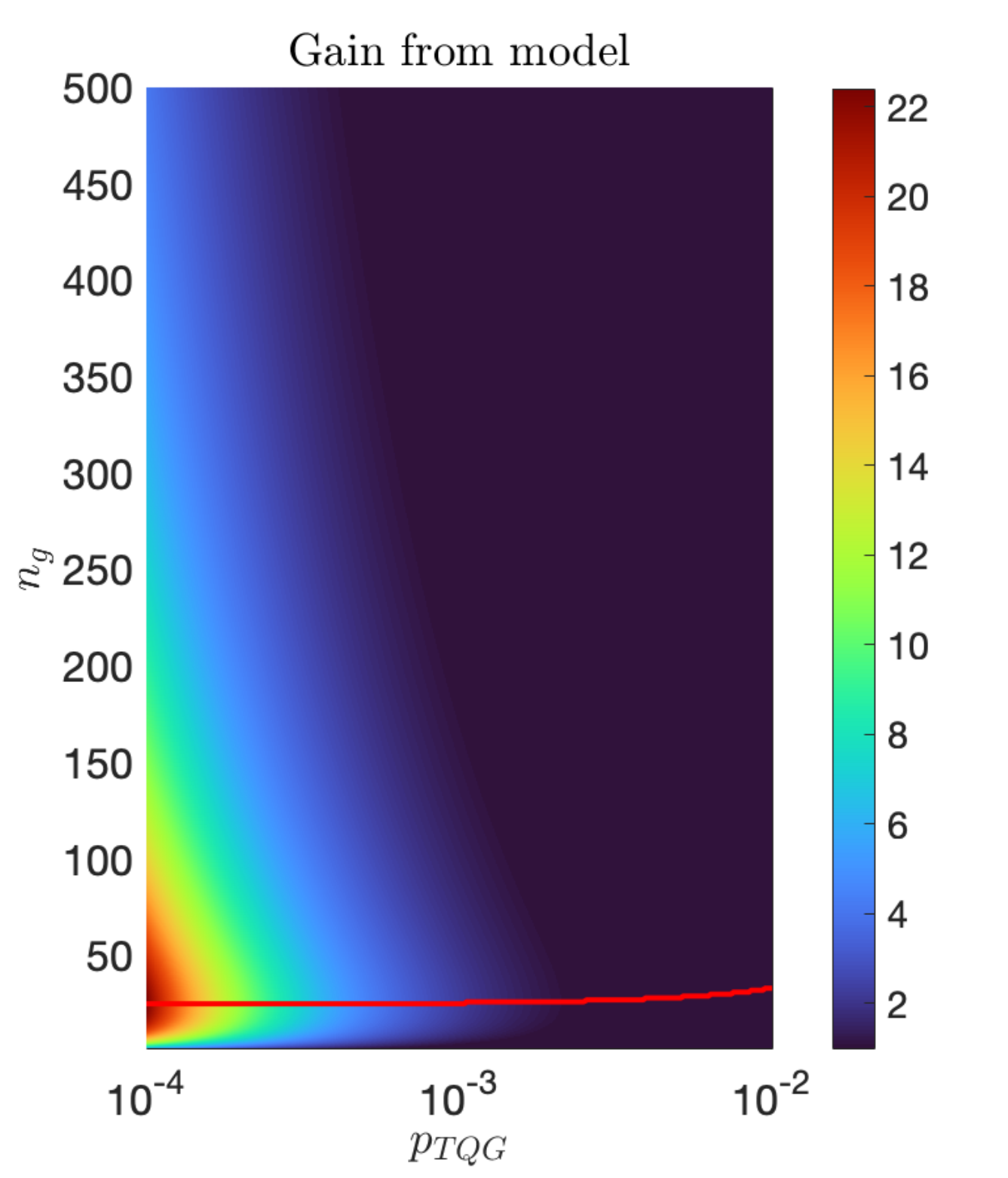}
    \caption{Gain, defined as the physical error when not using error correction divided by the logical error when using error correction calculated from eq. \eqref{eq:expression_gain} with $C=1$ and $N=75$. Here, we have assumed that all operations have the same error rate, $p_\mathrm{TQG}$. The number of logical gates performed before error correction, as well as the number of physical gates, is given by $n_g$. The red line marks the value of $n_g$ which maximizes the gain for every given $p_\mathrm{TQG}$. As is shown in eq. \eqref{eq:expression_gain}, below the threshold the maximum gain is achieved for a constant value of $n_g$, independent of $p_\mathrm{TQG}$.}
    \label{fig:modelgain}
\end{figure}
We can put this in more mathematical terms; Since all codes we study correct one error, the dominant term in the logical error rate is proportional to the probability of getting a weight two error, i.e. an error acting on two data qubits. We use proportional here, since some weight two errors are equal to a weight one error modulo one of the code stabilizers. Call the logical error rate $\epsilon_L$, and the number of operations used for error correction $N$. Assuming that all operations have an error rate of $p$ and neglecting higher order terms, we then have 
\begin{figure*}
    \includegraphics[width = \textwidth]{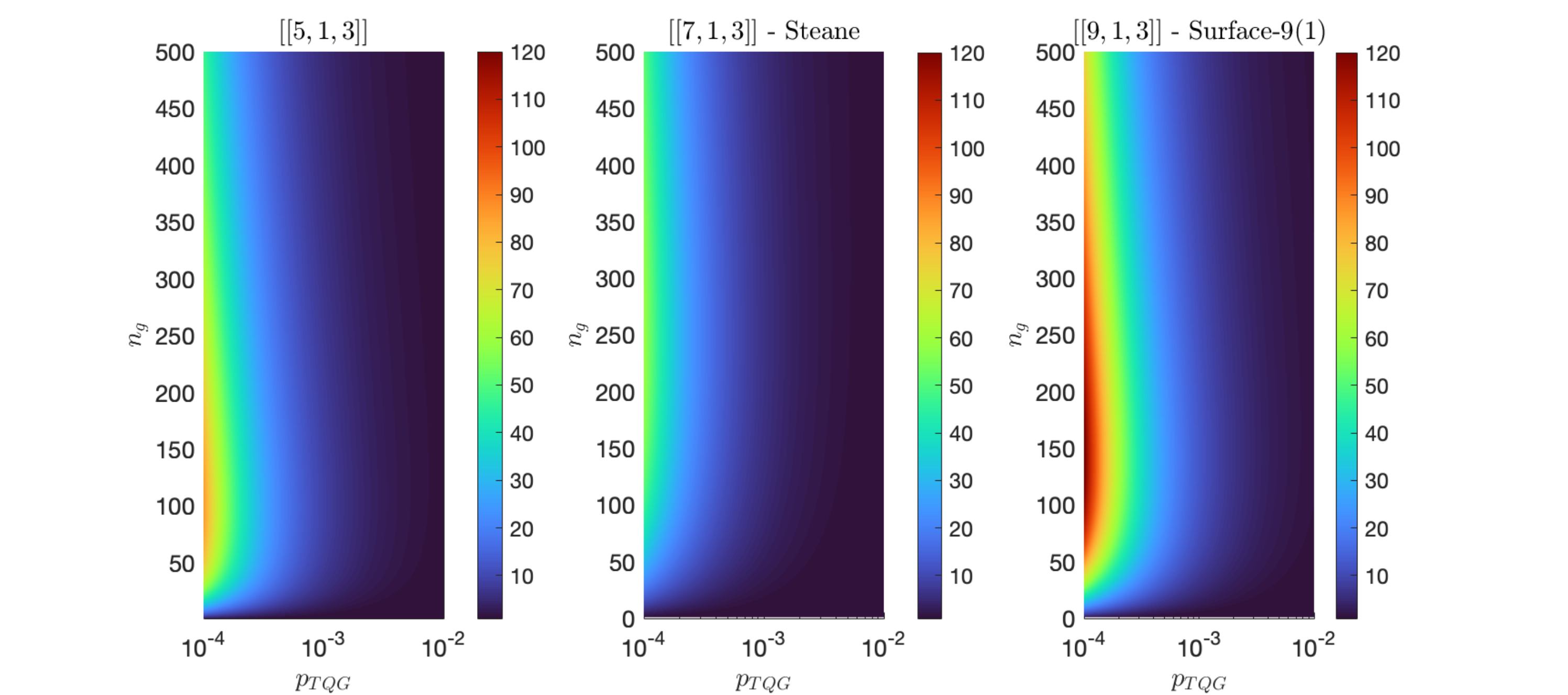}
    \caption{Gain of the logical circuit with error correction compared to the physical circuit. The $\llbracket5,1,3\rrbracket$ code achieves maximum gain at $n_g = 99$, the Steane code at $n_g = 244$ and Surface-9(1) at $n_g=140$.}
    \label{fig:gain}
\end{figure*}
\begin{align}
    \label{eq:approxerrorrate}
    \epsilon_L \propto p^2\binom{N}{2}.
\end{align}
Adding $n_g$ logical Pauli gates before error correction, and assuming that the physical operations again have error rate $p$, we get
\begin{align}
    \label{eq:tot_errorrate}
    \epsilon_L \propto p^2\binom{N + 3n_g}{2} = \frac{p^2}{2}(N + 3n_g)(N + 3n_g - 1),
\end{align}
since a logical Pauli gate on a distance three code corresponds to 3 physical Pauli gates. In the physical case however, the total error rate $\epsilon$ after $n_g$ operations, again neglecting higher order terms, is
\begin{align}
    \epsilon \propto p\binom{n_g}{1} = pn_g.
\end{align}
Thus, the gain $g$ can be written
\begin{align}
    \label{eq:expression_gain}
    g = \frac{\epsilon}{\epsilon_L} = \frac{C}{p} \cdot \frac{n_g}{(N + 3n_g)(N + 3n_g - 1)},
\end{align}
where $C$ is a constant determined by the proportion of weight two errors that are not equivalent to a weight one error modulo a stabilizer. With the restriction that $n_g \geq 0$ and $N >> 1$ for error correction, the gain function is concave and we can conclude that it has a maximum. Moreover, we can conclude that below the threshold, the gain is maximized for a constant value of $n_g$ independent of $p_\mathrm{TQG}$.

Figure \ref{fig:modelgain} shows a color map plot of a modified version of eq.\eqref{eq:expression_gain}, with a second order term added to the numerator, against both $n_g$ and $p$, with $N=75$ and $C = 1$. The addition of the second order term captures the behavior of the gain function for  values of $p$ above the threshold, where no gain from error correction can be expected and the optimal value of $n_g$ would therefore be one which gives a completely mixed state. In this model, the gain can be seen to achieve a maximum at $n_g=25$, i.e. $N/3$ for values of $p$ below the threshold. This means that in this simplified model where all operations have equal error rates, the optimal value of $n_g$ is the one where the number of physical gates used before error correction is $N/3$. This means that the optimal number of operations used before error correction is the same as the number of operations used in error correction. Moreover, the gain increases by one order of magnitude when there is a reduction in $p$ of one order of magnitude. This agrees well with what we would expect, as the error of a distance 3 code should scale as $p_\mathrm{TQG}^2$.

Our simple model can be compared to the full simulations of the different codes, which can be seen in Figure \ref{fig:gain} where we plot the gain as a color map against both the gate number $n_g$ and the error rate $p_{\mathrm{err}}$. The qualitative behavior is very similar to that of figure \ref{fig:modelgain}, with a constant maximum for the gain below the threshold error rate. It should be noted that here, the optimal $n_g$ is not just the one which corresponds to the number of operations used in the error correction protocol. This is because different types of operations, e.g. TQG and SQG, are assigned different error rates. 

When it comes to the performance of the codes, it is clear that both the $\llbracket5,1,3\rrbracket$ and Surface-9(1) codes outperform the Steane code with a large margin, both in terms of the threshold error and the maximum gain. The $\llbracket5,1,3\rrbracket$ and Surface-9(1) codes have very similar threshold values, as was established in Sec. \ref{sec:PseudoThresh}, but the maximum gain of Surface-9(1) is higher. 

With eq. \eqref{eq:expression_gain} in mind, we can say something about why these differences occur. We can see that the characteristics of a QEC code are determined by $N$, $C$ and also, in our case, the share of single-qubit operations of $N$. The Steane code in our case has a high maximum $N$ compared to the other two, and the $C$ apparently does not compensate for this. In the case of Surface-9(1), it is clear that the value of $C$ compensates for the higher number of operations it uses compared to the $\llbracket5,1,3\rrbracket$ code. This is likely also the reason why Surface-9(1) achieves a higher gain than the $\llbracket5,1,3\rrbracket$ code: If more weight two errors are correctable by the code, adding more error locations in the form of physical gates should have less of an impact. 

One thing to take into account when running experiments is the fact that the gain decreases slowly for values of $n_g$ above the optimal value. This means that the error correction overhead might be reduced by increasing $n_g$ above the optimal value, for only a small cost in decreased gain. This depends on the SQG error however, as the gain will decrease faster for higher SQG errors. This can be seen in figure \ref{fig:modelgain}, where all errors are the same. Assuming an experimental RE system can reach a value of $p_\mathrm{TQG}=10^{-3}$, the possible gain in error per gate is approximately 20 for both $\llbracket5,1,3\rrbracket$ code and Surface-9(1), and approximately 10 for the Steane code.



%% file: Sections/Conclusions.tex
We have studied the performance of three QEC codes, namely the minimal $\llbracket5,1,3\rrbracket$ code, the $\llbracket7,1,3\rrbracket$ Steane code, and the $\llbracket9,1,3\rrbracket$ Surface-9(1), using density matrix simulations, in the context of Rare Earth quantum computing. We required completely sequential gate operations, and found that the values of the pseudothresholds of all codes are highly dependent on whether resting errors are included or not. It was found that with resting errors, none of the codes would allow an RE quantum computer, using currently expected gate fidelities, to stay below the pseudothresholds required for fault tolerant operation. However, when resting errors can be neglected due to long ground state $T_2$, beneficial QEC would be possible. This is feasible to achieve experimentally. We also found that even when requiring sequential operations, the magnitude of resting errors can be reduced to a point where they are negligible in comparison to other error sources by extending the spin $T_2$; For the codes and protocols we studied, it is sufficient that the spin $T_2$ is around two orders of magnitude larger than the optical $T_2$, however a longer spin $T_2$ is required for higher distance codes.

We also compared the error per gate when using QEC to a physical circuit. We found that the optimal number of gates is determined by the number of operations used by the QEC protocol. The Steane code was found to be inferior to the other options both in terms of its pseudothreshold and in terms of the improvement in error rate compared to a physical circuit. The $\llbracket5,1,3\rrbracket$ code and Surface-9(1) were found to have very similar pseudothresholds of around $p_\mathrm{TQG}=4\cdot10^{-3}$ and $p_\mathrm{TQG}=3\cdot10^{-3}$ respectively. In terms of gain compared to a physical circuit, Surface-9(1) was found to be slightly better than the $\llbracket5,1,3\rrbracket$ code but at $p_\mathrm{TQG}=10^{-3}$ both had gain values of around 20, compared to a gain value of around 10 for the Steane code. Given the constraints of working in the NISQ regime, an experimentalist would likely be better off demonstrating QEC with the $\llbracket5,1,3\rrbracket$ code rather than Surface-9(1) in an early phase. This is because the $\llbracket5,1,3\rrbracket$ code uses fewer qubits, yet still has a slightly higher threshold error. As the number of available qubits increases, the surface codes become more attractive. This is partly due to their higher efficiency, but also the straightforward way in which higher distance codes can be implemented.

%% file: Sections/Background_Theory.tex
\subsection{\label{ssec:QEC}Quantum Error Correction}
In this section we give an overview of QEC that is relevant for this work. 

\subsubsection{\label{sssec:StabliserCodes}Stabilizer Codes}

All of the QEC codes used in this work belong to a class of codes known as \textit{Stabilizer Codes}. These can be compactly described in terms of a number of commuting operators belonging to the Pauli group, consisting of tensor products of the Pauli operators $X$, $Y$ and $Z$ together with the identity $I$ and multiplicative factors $\pm 1$ and $\pm i$ \cite{nielsen2010quantum}. These operators are known as the \textit{Stabilizer} of the code \cite{nielsen2010quantum, Devitt_2013}. Logical code words $\ket{0_L}$ and $\ket{1_L}$ are defined as simultaneous $+1$ eigenstates of all of the operators in the stabilizer \cite{nielsen2010quantum,Devitt_2013, Knill2005}. Errors can be detected and characterized through parity measurements of the stabilizers, and based on the measured \textit{error syndrome} $S_i$, a correction operator can be applied \cite{gottesman2009introduction}. The error syndrome consists of the recorded parities of all of the stabilizer measurements, and can be seen as a binary word, e.g. $S_{i} = \begin{pmatrix}1 & 1 & -1\end{pmatrix}\leftrightarrow \begin{pmatrix}0 & 0 & 1\end{pmatrix}$ for a hypothetical code with only three stabilizers. The error syndromes can be conveniently labeled by the corresponding decimal number, so that e.g. $S_3 = \begin{pmatrix} 0 & 1 & 1 \end{pmatrix}$.
\\

Quantum error correcting codes can also be described on the form $\llbracket n,k,d\rrbracket$, where $n$ is the number of physical qubits used for encoding, $k$ is the number of encoded qubits and $d$ is called the \textit{distance} of the code. The distance $d$ is a measure of how different the code words of a given code are. The distance determines how many errors that a code can correct, since a distance of $d=2t+1$ is required to correct $t$ errors. The distance can be seen as the minimum weight of any logical operator on the code, that is, the minimum number of qubits on which a logical operator on the code differs from the identity. Logical operators for stabilizer codes can be defined by picking operators which commute with all members of the stabilizer group, and which obey the commutation relations of the equivalent physical operators. One of the most famous examples of quantum error correcting codes, the Steane code, is a $\llbracket 7,1,3 \rrbracket$ code, which means it uses 7 qubits to encode one logical qubit with distance three \cite{nielsen2010quantum,Devitt_2013,gottesman2009introduction}.\\ 
\\

\subsection{\label{sec:errormodel}Error Models}
In this section we describe the error model used throughout the simulated experiment.

\subsubsection{\label{ssec:GateErrors}Single and Two Qubit Gate Errors}
We have modeled single qubit gate (SQG) errors by assuming a perfect gate is followed by an error process $\mathcal{E}_{SQG}$, given by 

\begin{eqnarray}
\label{eq:SQGErrors}
    \mathcal{E}_{SQG}(\rho) = (1 - p_{X} - p_{Y}  - p_{Z})\rho \\ \nonumber+ p_{X}X\rho X^\dagger + p_{Y}Y\rho Y^\dagger + p_{Z}Z\rho Z^\dagger,
\end{eqnarray}
where $I$, $X$, $Y$ and $Z$ are the Pauli operators with corresponding Pauli matrices $\sigma_I, \ \sigma_X, \ \sigma_Y, \ \sigma_Z$. Typically, we set a value $p_{\mathrm{SQG}}$ for the total SQG error rate and use $p_X = p_Y = p_Z = p_{\mathrm{SQG}}/3$, i.e. the standard single qubit depolarizing channel \cite{Knill2005,nielsen2010quantum}. For two qubit gates (TQG), such as the controlled-NOT, we assume that the error operators are given by the non trivial two qubit Pauli group operators, i.e. tensor products on the form $\sigma_i\otimes\sigma_j$ with corresponding probabilities given by $p_{ij}$, with $i,j$ being $I$, $X$, $Y$ or $Z$. The error process $\mathcal{E}_{TQG}$ for TQG is given by 

\begin{align}
\label{eq:TQGErrors}
    \mathcal{E}_{TQG}(\rho) = (1 - \sum_{ij\neq II}p_{ij})\rho + \sum_{ij\neq II}p_{ij}\sigma_i \otimes \sigma_j \rho \sigma_i^\dagger \otimes \sigma_j^\dagger.
\end{align}
For TQG we typically set an error rate $p_{\mathrm{TQG}}$ and use $p_{ij \neq II} = p_{\mathrm{TQG}}/15$ corresponding to a two qubit depolarizing channel \cite{Knill2005}. 
The full operation for an arbitrary gate $\mathcal{G}$ is thus given by 
\begin{align}
    G(\rho) = \mathcal{E}(\sigma_G\rho \sigma_G^\dagger),
\end{align}
where $\sigma_G$ is the operation matrix for the error free $G$.

\subsubsection{\label{ssec:Idleerrors}Idle Qubit Errors}
Because the ground state life time of the rare earths can be days, amplitude damping is neglected on resting qubits. Phase damping is considered, and since operations are assumed to be sequential, it is applied on all qubits except the involved qubits of a gate during gate operations, where instead normal gate errors are applied. Phase damping is also applied to all qubits except the target during initialisation and readout. 

The standard form of the phase damping channel for one qubit is given by the operation matrices
\begin{align}
    E_0 &= 
    \begin{pmatrix}
    1 & 0 \\
    0 & \sqrt{1-\gamma}
    \end{pmatrix} \\
    E_1 &= 
    \begin{pmatrix}
    0 & 0 \\
    0 & \sqrt{\gamma}
    \end{pmatrix}
\end{align}
\cite{nielsen2010quantum}. However, the channel can be equivalently represented by the operation matrices
\begin{align}
    \tilde{E_0} &= \sqrt{\alpha}
    \begin{pmatrix}
    1 & 0 \\
    0 & 1
    \end{pmatrix} \\
    \tilde{E_1} &= \sqrt{1 - \alpha}
    \begin{pmatrix}
    1 & 0 \\
    0 & -1
    \end{pmatrix},
\end{align}
where $\alpha = (1 + \sqrt{1-\gamma})/2$ \cite{nielsen2010quantum}. We use the latter of these representations, since it can save some calculations during simulations. In the case where the groundstate coherence time $T_\mathrm{2, spin}$ is used to calculate the damping coefficient, we use
\begin{align}
    \gamma = 1 - e^{-t/2T_\mathrm{2, spin}},
\end{align}
where $t$ is the idle time given by the duration of the operation. The different operations used are SQG, TQG, initialisation and readout, and we denote their operation times by $t_\mathrm{SQG}$, $t_\mathrm{TQG}$, $t_\mathrm{init}$ and $t_\mathrm{RO}$ respectively. 

\subsubsection{\label{ssec:init&RoErrors}Initialisation and Readout Errors}
Qubits are for simplicity always initialised in the $\ket{0}$ state, and it is assumed that there is a probability, $p_\mathrm{init}$, of an initialisation error when it is instead initialised as $\ket{1}$. The density matrix $\rho$ after initialisation for a single qubit with initialisation error rate $p_\mathrm{init}$ is thus
\begin{align}
    \rho = (1-p_\mathrm{init})\ket{0}\bra{0} + p_\mathrm{init}\ket{1}\bra{1}:=
    \begin{pmatrix}
    1-p_\mathrm{init} & 0 \\
    0 & p_\mathrm{init}
    \end{pmatrix}.
\end{align}
We assume that readout in the computational basis ${\ket{0}, \ket{1}}$ is subject to a declaration error, e.g. where a measurement result of $-1$ is incorrectly declared as $+1$ and vice versa. In the general case we use a symmetric declaration error, where the states $\rho_\pm$ after measurement of $+1$ and $-1$ are given by 
\begin{align}
    \rho_+ &= (1-p_\mathrm{RO})\cdot p_+ \cdot\rho_{+1} + p_\mathrm{RO}\cdot p_- \rho_{-1}, \\
    \rho_- &= (1-p_\mathrm{RO})\cdot p_- \cdot\rho_{-1} + p_\mathrm{RO}\cdot p_+ \rho_{+1},
\end{align}
where $\rho_{\pm1}$ are the resulting density matrices after error free measurement, $p_\pm$ is the probability to measure $+1$ and $-1$ and $p_\mathrm{RO}$ is the probability of a declaration error. An idiosyncrasy of the RE system is that declaration errors are asymmetrical, so that a measurement result of $-1$ has a probability to be declared as $+1$, but a declaration of $-1$ is always correct. This is due to the nature of the readout via the dipole mechanism where only one of the states is excited \cite{Walther2015}. The states $\rho_\pm$ after measuring $+1$ and $-1$ are given by
\begin{align}
    \rho_+ &= p_+\cdot\rho_{+1} + p_{RO}\cdot p_-\rho_{-1},\\
    \rho_- &= (1-p_{RO})\cdot p_-\cdot\rho_{-1} 
\end{align}
in the asymmetrical case.

%% file: Sections/DensityMatrixSim.tex
The density matrix simulations were performed using a Matlab framework. The density matrix approach was chosen for its ability to model arbitrary quantum processes, such as, for example, $T_2$ decoherence. It would also be possible in future implementations to find exact representations for all gate operations, including different error sources. In contrast to the more common Monte Carlo based simulations used in e.g. \cite{Tomita2014Surface17, Chao2018, Bermudez2017}, a completely analytical method was chosen. Faulty gates are modeled as a perfect gate operation given by e.g. one of the Pauli operators, followed by a noise channel as described in appendix. \ref{ssec:GateErrors}. Measurements in the $Z$-basis are also handled analytically, where the resulting density matrix is calculated for each measurement outcome and then handled separately. With this method, simulating one round of error correction is done by calculating the final states $\rho_{i}$ corresponding to error syndromes $S_i$, e.g. $\begin{pmatrix} 0 &0 &1\end{pmatrix}$, and then summing all of these weighted by the probability $p_i$ of measuring syndrome $S_i$ as 
\begin{equation}
    \rho_f = \sum_ip_i\rho_i,
\end{equation}
where $\rho_f$ is the final density matrix. This method is convenient for smaller codes, as it does not involve simulating a large number of rounds to get a large enough data set. However, for larger codes with more stabilizer measurements, the problem quickly becomes prohibitively complex and expensive to calculate, even when several optimization techniques such as sparse matrix operations and caching of expensive function calls are employed. The method lends itself well to parallelisation since all outcomes are calculated independently, so it is possible that considerable speedup could be achieved using a GPU. 


    
        
        
        

\subsection{\label{sec:PerformanceMeasure}Calculating Logical Error Rate and Code Performance}
For our code performance measure, we have used the logical error rate defined as the probability of an uncorrectable error being present after the error correction procedure. After an imperfect round of error correction, the density matrix will be in a mixed state. This can be taken to a pure state by performing one round of perfect, i.e. error free, error correction. This will take correctable errors to the correct codeword, and uncorrectable errors to the opposite code word. To calculate logical error rates, we define the \textit{reduced logical density matrix} $\rho^{red}_L$ for pure states by
\begin{align}
    \rho^{red}_L = \mathbf{tr}(\rho')I + \mathbf{tr}(X_L\rho')X + \mathbf{tr}(Y_L\rho')Y + \mathbf{tr}(Z_L\rho')Z, \label{eq:rhoL}
\end{align}
where $X_L$, $Y_L$ and $Z_L$ are the logical Pauli operators defined for the given code, and $X$, $Y$ and $Z$ are the regular Pauli matrices. By writing the error free state $\ket{\psi_L}$ of the logical qubit as a linear combination of the logical basis vectors $\ket{0_L}$, $\ket{1_L}$ by
\begin{align}
    \ket{\psi_L} = \alpha \ket{0_L} + \beta \ket{1_L},
\end{align}
we can define the \textit{reduced logical state vector} $\ket{\psi^{red}_L}$ as
\begin{align}
    \ket{\psi^{red}_L} = \alpha\ket{0} + \beta\ket{1} \label{eq:psiL}.
\end{align}
The logical error rate $\epsilon_L$ can then be calculated from the fidelity between $\rho^{red}_L$ and $\ket{\psi^{red}_L}$ by

\begin{align}
    \epsilon_L = 1 - \bra{\psi^{red}_L}\rho^{red}_L\ket{\psi^{red}_L}.
\end{align}
